\documentclass[aps,preprint,nofootinbib,superscriptaddress,prc]{revtex4}
\usepackage{amssymb}

\usepackage{epsfig}
\usepackage[bookmarksnumbered,bookmarksopen,colorlinks,citecolor=blue,linkcolor=blue] {hyperref}
\begin{document}

\date{\today}

\title{Extraction of symmetry energy coefficients from the mass differences of isobaric nuclei }

\author{Junlong Tian}
 \email{tianjunlong@gmail.com}
 \affiliation{ School of Physics and Electrical Engineering,
Anyang Normal University, Anyang 455000, People's Republic of
China }

\author{Haitao Cui}
\affiliation{  School of Physics and Electrical Engineering,
Anyang Normal University, Anyang 455000, People's Republic of
China }

\author{Kuankuan Zheng}
\affiliation{  School of Physics and Electrical Engineering,
Anyang Normal University, Anyang 455000, People's Republic of
China }

\author{Ning Wang}
\affiliation{ Department of Physics, Guangxi Normal University,
Guilin 541004, People's Republic of China }

\begin{abstract}
The nuclear symmetry energy coefficients of finite nuclei are extracted by using the differences between the masses of isobaric nuclei. Based on the masses of more than 2400  nuclei with $A=9-270$, we investigate the model dependence in the extraction of symmetry energy coefficient. We find that the extraction of the symmetry energy coefficients is strongly correlated with the forms of the Coulomb energy and the mass dependence of the symmetry energy coefficient adopted. The values of the extracted symmetry energy coefficients increase by about 2 MeV for heavy nuclei when the Coulomb correction term is involved. We obtain the bulk symmetry energy coefficient $S_0=28.26\pm1.3$ MeV and the surface-to-volume ratio $\kappa=1.26\pm 0.25 $ MeV if assuming the mass dependence of symmetry energy coefficient $a_{\rm sym}(A)=S_0(1-\kappa/A^{1/3})$, and $S_0=32.80\pm1.7$ MeV, $\kappa=2.82\pm0.57$ MeV when $a_{\rm sym}(A)=S_0 (1+\kappa/A^{1/3})^{-1}$ is adopted.
\end{abstract}

\maketitle

\section{\label{sec:level2}Introduction}
The symmetry energy coefficient plays a key role, not only in nuclear physics,
such as the dynamics of heavy-ion collisions induced by radioactive beams and the structure of exotic nuclei near the nuclear drip lines \cite{Dani02,Stein05,Bara05,Latt07,liba08,Dong11},
but also a number of important issues in astrophysics, such as the dynamical evolution of the core collapse of a massive star and the associated explosive nucleosynthesis \cite{Latt00,Horo01,Todd05,Shar09,Kumar11,tianwd11,Fatto13}.
In the global fitting of the nuclear masses in the framework
of the liquid-drop mass formula, the symmetry coefficient
$a_{sym}$ of finite nuclei enters as a mass-dependent phenomenological parameter \cite{Janec03,Ono04,WangN2,Oyam10,Nik11}.
In the symmetry energy coefficient $a_{sym}$, the volume coefficient $S_0$ which represents the nuclear symmetry energy at normal density and the surface coefficient (or the surface-to-volume ratio $\kappa$) are two important quantities. In the realistic calculations of nuclear masses, two different forms for description of the mass dependence of $a_{sym}$ are frequently used. One is $a_{\rm sym}(A)=S_0(1-\kappa/A^{1/3})$ \cite{Moller95,Stoi07,Satula06,Samad07,Kolom10,meihua12,macw12,Jiangh12}, the other is $a_{\rm sym}(A)=S_0 (1+\kappa/A^{1/3})^{-1}$ \cite{Dani03,Liu10,Die07,Dong13}.
However, the values of the parameters $S_0$ and $\kappa$ are quite different in different theoretical frameworks. It is therefore necessary to investigate the influence of model dependence on the extraction of nuclear symmetry coefficient.

Nuclear mass is one of the most precisely experimentally
determined quantity in nuclear physics. It can provide
information of the symmetry energy coefficient $a_{sym}(A)$ through the liquid-drop mass systematics.
In the Bethe-Weiszacker (BW) mass formula \cite{Weiz35,Bethe36}, the binding energy of a nucleus with the mass number $A$, the charge $Z$ and the neutron number $N$,
is expressed as
\begin{eqnarray}
B(A,Z)=a_{v}A-a_{s}A^{2/3}-a_{c}\frac{Z^{2}}{A^{1/3}}-a_{sym}\frac{(N-Z)^{2}}{A}+\delta,
\end{eqnarray}
with
\begin{eqnarray}
\delta=\pm a_{p}A^{-1/2} ~ or ~ 0,
\end{eqnarray}
where the ``+" is for even-even nuclides, the ``--" is for odd-odd nuclides,
and for odd-A nuclides (i.e. even-odd and odd-even) $\delta=0$.
The a$_{v}$, a$_{s}$, a$_{c}$, a$_{sym}$ and a$_{p}$ are the volume, surface, Coulomb, symmetry and pairing energy coefficients, respectively.

Based on the BW mass formula Eq.(1), the binding energy difference between two isobaric nuclei with $\Delta Z$, which is a multiple of 2, is written as,
\begin{eqnarray}
B(A,Z+1)-B(A,Z-1)=[8a_{sym}\frac{(A-2Z)}{A}-4a_{c}\frac{Z}{A^{1/3}}],
\end{eqnarray}
\begin{eqnarray}
B(A,Z+2)-B(A,Z-2)=2[8a_{sym}\frac{(A-2Z)}{A}-4a_{c}\frac{Z}{A^{1/3}})],
\end{eqnarray}
\begin{eqnarray}
B(A,Z+3)-B(A,Z-3)=3[8a_{sym}\frac{(A-2Z)}{A}-4a_{c}\frac{Z}{A^{1/3}}],\\
\nonumber ......
\end{eqnarray}
\begin{eqnarray}
B(A,Z+n)-B(A,Z-n)=n[8a_{sym}\frac{(A-2Z)}{A}-4a_{c}\frac{Z}{A^{1/3}}],
\end{eqnarray}

From the Eqs.(3)-(6), we can obtain the following expression,
\begin{eqnarray}
a_{sym}(A)=\frac{A}{8(A-2Z)}[\frac{B(A,Z+i)-B(A,Z-i)}{i}+4a_{c}\frac{Z}{A^{1/3}}],
\end{eqnarray}
where $i=1, 2, 3,...,n$, and $n$ is the count of isobaric nuclei pairs for a given mass number $A$.
From Eq.(7), we can see that the volume, surface and pairing terms are canceled each other from the difference of two isobaric nuclei with $\Delta Z$ is a multiple of 2, the symmetry energy coefficient $a_{sym}(A)$ depends on the number $i$, the Coulomb energy coefficient $a_{c}$ and the chosen central nucleus ($A,Z$).
But through the summation of Eqs.(3)-(6), the effect depending on $i$ is canceled. we obtain the following expression,
\begin{eqnarray}
\sum_{i=1}^{n}[B(A,Z+i)-B(A,Z-i)]=\frac{n(n+1)}{2}[8a_{sym}\frac{(A-2Z)}{A}-4a_{c}\frac{Z}{A^{1/3}}].
\end{eqnarray}
Then the symmetry energy coefficient can be extracted,
\begin{eqnarray}
a_{sym}(A)=\frac{A}{4(A-2Z)}\{\frac{1}{n(n+1)}\sum_{i=1}^{n}[B(A,Z+i)-B(A,Z-i)]+2a_{c}\frac{Z}{A^{1/3}}\}.
\end{eqnarray}
Eq.(9) is the average value of Eq.(7) with $i$ from $1$ to $n$ for a given mass number $A$.
The central reference nucleus ($A,Z$) is usually selected according to the following procedure.
We assume there is $k$ isobaric nuclei for a given mass number $A$, $Z_{min}$ and $Z_{max}$ denote the minimum and the maximum charge numbers.
if $k$ is odd number $n=(k-1)/2$, and then the central reference nucleus ($A,\frac{Z_{max}+Z_{min}}{2}$) is selected.
if $k$ is even number then $n=\frac{k}{2}-1$, the central reference nucleus ($A,Z_{min}+n$) or ($A,Z_{max}-n$) is selected,
and finally we take the average value of these two cases in calculation $a_{sym}(A)$ by using Eq. (9). In this work, the symmetric nucleus ($N=Z$) is not chosen as the central nucleus, and the symmetric nucleus does not enter the calculation in Eq.(7) and Eq.(9).

On the other hand, the liquid drop energy of a nucleus B(A,Z) can be expressed as
\begin{eqnarray}
B(A,Z)=B^{exp}(A,Z)-E_{sh}(A,Z)-E_{W}(A,Z),
\end{eqnarray}
where $B^{exp}(A,Z)$ is the experimental measured nuclear binding energy compiled in Ref. \cite{AME2012},
$E_{sh}(A,Z)$ and $E_{W}(A,Z)$ denote the shell correction and the Wigner energy, respectively. The shell correction energy is selected from the KTUY \cite{KTUY05} model,
which are global nuclear mass model with a high accuracy and good extrapolation. For the Wigner energy, we take the form $E_{W}=10exp(-4.2|I|)$ as in Ref. \cite{Liu10, Myer97}, where $I=(N-Z)/A$.
So long as the Coulomb energy expression and its coefficients are determined, the symmetry energy coefficient can be calculated by using Eqs.(9) and (10).¡¡

The paper is organized as follows. In Sec.II, the Coulomb energy expression and its coefficients are determined from the difference of the experimental binding energies for 88 pairs of mirror nuclei in the region $11 \leq A \leq 75 $. In Sec.III, we extract the average symmetry energy coefficient $a_{sym}(A)$ by using the differences between the masses of isobaric nuclei, and we obtain the values of $S_{0}$ and  $\kappa$ by performing a two-parameter fitting to $a_{sym}(A)$. The effect of Coulomb energy term and the shell correction energy on the symmetry energy coefficient is studied in Sec.III. Finally a summary is given in Sec. IV.

\section{\label{sec:level1}Coulomb energy coefficients}
Eqs. (7) and (9) are obtained by selecting the Coulomb energy expression $E_{c}=\frac{a_{c}}{A^{1/3}}Z^{2}$ (set I, see Table I ). The Coulomb energy coefficient $a_c$ is determined from the difference of the experimental binding energies for 88 pairs of mirror nuclei in the region $11 \leq A \leq 75 $, which are found in the 2012 Atomic mass Evaluation (AME2012) \cite{AME2012}. There are no mirror nuclides with $ A > 75$. This method is used in Ref. \cite{Kirson08}.
It is well known that mirror nuclei are pairs of nuclei with same mass number $A$, but with $Z_{1}=N_{2}$ and $Z_{2}=N_{1}$, \emph{i.e.} with neutrons and protons interchanged. Given charge independence of the nuclear force, the binding energies of mirror pairs can differ only in their Coulomb energies. The difference in the binding energy between two mirror nuclei is thus $\Delta B=\Delta E_{c}=a_c(N^{2}-Z^{2})/A^{1/3}=a_c\Delta ZA^{2/3}$, where $\Delta Z$ is the difference in proton number between the two mirror nuclei. The quantity $\Delta B/\Delta Z$ should be linear in $A^{2/3}$ and pass through the origin. The slope of the line is the empirical coefficient $a_c$ of the Coulomb energy term.

In Fig.1 (a) the binding energy differences of the 88 pairs of mirror nuclei, scaled by charge difference $\Delta Z$ are plotted against $A^{2/3}$ and are seen to lie on a straight line. The value of $\Delta Z$ rages from 1 (32 cases) to 5 (1 case). If we adopt the Coulomb energy expression set I, the least squares fit gives a straight line (dashed line) and pass through the origin, with the slop $a_c\approx 0.625$ MeV and a root-mean-squared deviation (rmsd) of 336 keV. The red solid line is the best fit straight line with a rmsd of 121 keV, which has a slop of $0.715$ MeV, but the intercept is $-1.04$ MeV, a sizeable distance from the expected value of zero. It is implied that a physical effect has been overlooked. The missing term responsible for the non-zero intercepts is the contribution of charge exchange and all the other correction terms including the nuclear surface diffuseness correction. If we assume the Coulomb energy expression $E_{c}=\frac{a_{c}}{A^{1/3}}Z^{2}(1-bZ^{-2/3})$ (set II) by adding the Coulomb charge exchange and all the other correction terms, then $\Delta B/\Delta Z=a_c(A^{2/3}-\frac{2^{5/3}b}{3})$, which is compared with the fitting line $\Delta B/\Delta Z=0.715A^{2/3}-1.04$. The values of $a_{c}=0.715$ MeV and $b=1.374$ are obtained. If we put $b_{1}=\frac{5}{4}(\frac{3}{2\pi})^{2/3}\simeq0.764$, only the Coulomb charge exchange energy of the Fermi gas is included. Here $b=b_{1}+b_{2}=1.374$ is introduced to take into account the Coulomb charge exchange term ($b_{1}=0.764$), and all the other corrections including the nuclear surface diffuseness correction term ($b_{2}=0.610$).
\begin{figure}
\includegraphics[angle=-0,width= 1.0 \textwidth]{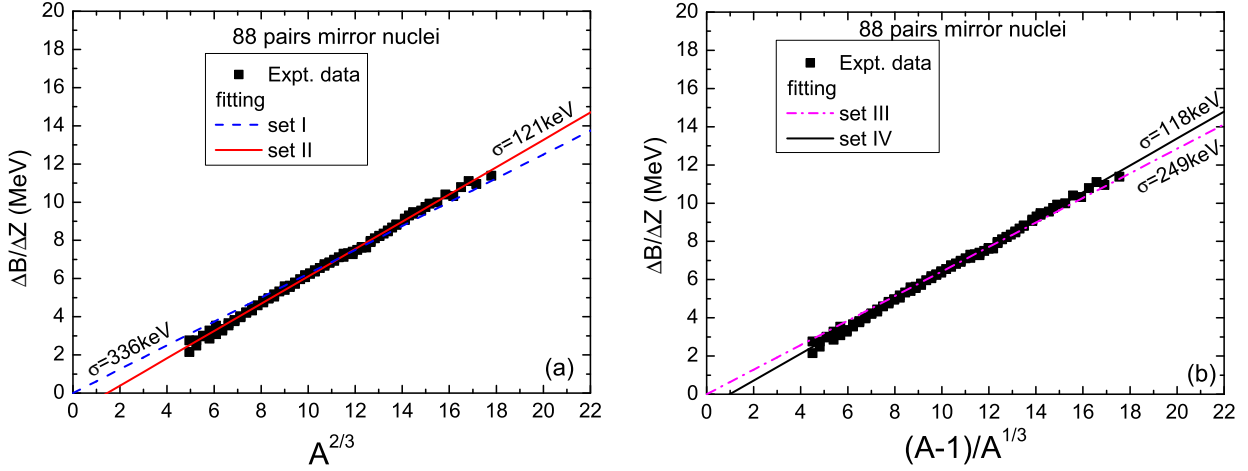}
 \caption{(Color online) Scaled 88 pairs mirror nuclei binding energy differences vs. $A^{2/3}$ (a) and vs. $(A-1)/A^{1/3}$ (b), and fitting lines for without take into account the Coulomb correction terms (dashed line and dash-dotted line) and with the correction terms (red solid line and black solid line).}
\end{figure}
\begin{table}
\caption{ The Coulomb energy coefficiens $a_{c}$ and $b$ are determined from the difference of the experimental binding energies for 88 pairs of mirror nuclei in the region $11 \leq A \leq 75 $ with the rmsd $\sigma$ for four sets Coulomb energy expressions.}
\begin{tabular}{ccccccc}
\hline\hline
                    &Coulomb energy $E_{c}$         &~~Mirror pairs $\frac{\Delta B}{\Delta Z}$~   & fitting $\frac{\Delta B}{\Delta Z}$ &~$a_{c}$(MeV)~  & ~$b$~ &~ $\sigma$(keV)~  \\ \hline
set I    & $a_{c}\frac{Z^{2}}{A^{1/3}}$    & $a_{c}A^{2/3}$  & $0.625A^{2/3}$  & 0.625        & $-$       & $336$   \\
set II   & $a_{c}\frac{Z^{2}}{A^{1/3}}(1-bZ^{-2/3})$ & $a_{c}(A^{2/3}-\frac{2^{5/3}b}{3})$ & $0.715A^{2/3}-1.04$  &  0.715    & 1.374 & 121 \\
set III  & $a_{c}\frac{Z(Z-1)}{A^{1/3}}$  & $a_{c}\frac{(A-1)}{A^{1/3}}$ & $0.642\frac{(A-1)}{A^{1/3}}$ &  0.642      & $-$    & 249 \\
set IV   & $a_{c}\frac{Z(Z-1)}{A^{1/3}}(1-bZ^{-2/3})$ & $a_{c}(\frac{(A-1)}{A^{1/3}}-\frac{2^{5/3}b}{3})$ & $0.704\frac{(A-1)}{A^{1/3}}-0.694$&  0.704   & 0.985  & 118 \\
 \hline\hline
\end{tabular}
\end{table}

In order to study the effect of the Coulomb energy on the symmetry energy coefficient, we change the Coulomb energy expression to $E_{c}=\frac{a_{c}}{A^{1/3}}Z(Z-1)$ (set III), because the Coulomb repulsion will only exit for more than one proton, $Z^{2}$ becomes $Z(Z-1)$. Then the Coulomb energy coefficient $a_c$ is determined following the above procedure. The difference in the binding energy between two mirror nuclei is $\Delta B=\Delta E_{c}=a_c[N(N-1)-Z(Z-1)]/A^{1/3}=a_c\Delta Z(A-1)/A^{1/3}$. The quantity $\Delta B/\Delta Z$ should be linear in $(A-1)/A^{1/3}$ and pass through the origin. Fig.1 (b) shows that the least squares fit gives $a_c\approx 0.642$ MeV (dash-dotted line), with a rmsd of 249 keV. While the black solid line is the best fit straight line with a rmsd of 118 keV, which has a slop of $0.704$ MeV and the intercept is $-0.694$ MeV not zero. If we assume the Coulomb energy expression $E_{c}=\frac{a_{c}}{A^{1/3}}Z(Z-1)(1-bZ^{-2/3})$ (set IV) by adding the Coulomb charge exchange and the other correction terms, then $\Delta B/\Delta Z\approx a_c(\frac{(A-1)}{A^{1/3}}-\frac{2^{5/3}b}{3})$, which is compared with the fitting line $\Delta B/\Delta Z=0.704\frac{(A-1)}{A^{1/3}}-0.694$. The values of $a_{c}=0.704$ MeV and $b=0.985$ are obtained

In Table I we list four sets Coulomb energy expressions $E_{c}$ and corresponding coefficients, the binding energy difference of mirror pairs $\frac{\Delta B}{\Delta Z}$ and its fitting result,  and the their rmsd $\sigma$ mentioned in Fig. 1. One sees that the rmsd for the Coulomb energy expressions
set I and set III are larger than those in expressions set II and set IV.
It is implied that the contribution of charge exchange and the other correction terms must be taken into account.
All four expressions in Table I, the set IV has the least rmsd of 118 keV for the binding energy differences of 88 pairs mirror nuclei.
So the Coulomb energy expression $E_{c}=\frac{0.704}{A^{1/3}}Z(Z-1)(1-0.985Z^{-2/3})$ (set IV) is adopted in the following calculations.

\section{\label{sec:level1}symmetry energy coefficient}

\begin{figure}
\includegraphics[angle=-0,width= 0.65\textwidth]{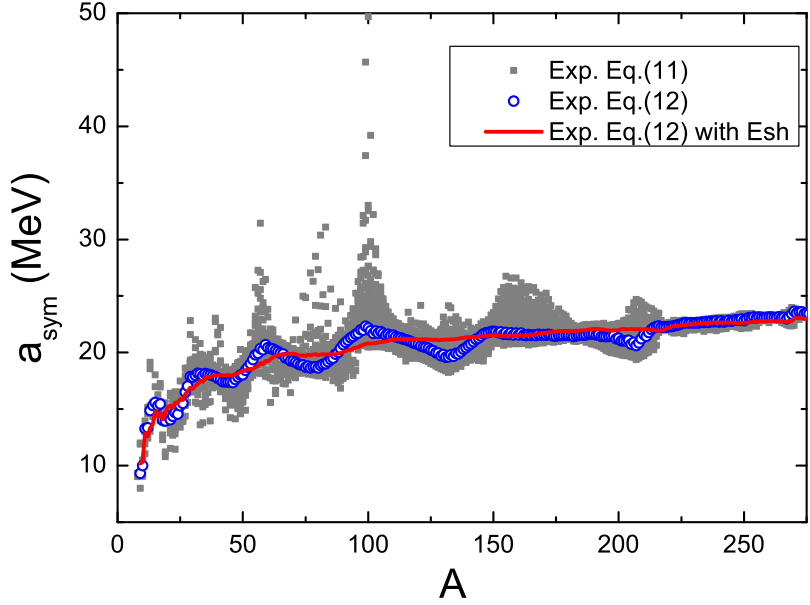}
 \caption{(Color online) Experimental symmetry energy coefficient $a_{sym}(A)$ as a function of mass number $A$ from Eq. (11) (solid squares) and Eq. (12) (open circles) without take into account the shell corrections (Esh), and with the shell corrections (red curve).}
\end{figure}
The Coulomb energy expression set I $E_{c}=a_{c}\frac{Z^{2}}{A^{1/3}}$ in the Eq.(1) is replaced by set IV $E_{c}=\frac{0.704}{A^{1/3}}Z(Z-1)(1-0.985Z^{-2/3})$, and we repeat the same procedure of extraction symmetry energy coefficient from Eqs.(3)-(9). we obtain the symmetry energy coefficient,
\begin{eqnarray}
\nonumber a_{sym}(A)=&&\frac{A}{8(A-2Z)}\{\frac{B(A,Z+i)-B(A,Z-i)}{i}\\
+&&\frac{0.704}{A^{1/3}}[4Z-2-0.985(\frac{8}{3}Z^{1/3}-\frac{2}{3}Z^{-2/3})]\},
\end{eqnarray}
and the average value is
\begin{eqnarray}
\nonumber a_{sym}(A)=&&\frac{A}{4(A-2Z)}\{\frac{1}{n(n+1)}\sum_{i=1}^{n}[B(A,Z+i)-B(A,Z-i)]\\
&&+\frac{0.704}{A^{1/3}}[2Z-1-0.985(\frac{4}{3}Z^{1/3}-\frac{1}{3}Z^{-2/3})]\}.
\end{eqnarray}

Figure 2 shows the experimental symmetry energy coefficient $a_{sym}(A)$ of nuclei as a function of mass number $A$. The solid squares denote the extracted symmetry energy coefficient from the Eq. (11). The reference nucleus ($A,Z$) is arbitrary known nucleus except for the symmetric nucleus, the liquid drop energy of $B(A,Z\pm i)$ is taken from the measured nuclear binding energy minus the Wigner energy. The open circles denote the extracted results from the Eq. (12), in which the central reference nucleus ($A,Z$) is selected. Insert the Eq.(10) into the Eq.(12), we can obtain the smooth symmetry energy coefficient. The thick red curve is the extracted experimental symmetry energy coefficient from the Eq. (12) by considering the shell correction energy of KTUY in Ref. \cite{KTUY05}. From Fig. 2 one can see that the existence of exceptionally large values of the symmetry energy coefficient at mass number $A\approx100$ that was also reported in Ref. \cite{Janec03}. The values of $a_{sym}(A)$ obtained in our approach by Eqs. (11) and (12) show some oscillations and fluctuations. When the shell corrections are taken into account, the fluctuations in the extracted $a_{sym}(A)$ are reduced effectively (thick red curve).

Eqs. (7) and (9) are obtained by selecting the Coulomb energy expression set I, while Eqs. (11) and (12) are obtained by selecting the Coulomb energy expression set IV in Table I.
In order to study the effect of the Coulomb energy on the symmetry energy coefficient, we change the Coulomb energy expression by selecting the expression set I-IV in Table I. In the same manner, if we take the Coulomb energy expression set II or III, the average symmetry energy coefficient can be extracted, respectively,
\begin{eqnarray}
\nonumber a_{sym}(A)=&&\frac{A}{4(A-2Z)}\{\frac{1}{n(n+1)}\sum_{i=1}^{n}[B(A,Z+i)-B(A,Z-i)]\\
&&+\frac{0.715}{A^{1/3}}(2Z-1.374\frac{4}{3}Z^{1/3})\},
\end{eqnarray}
or
\begin{eqnarray}
a_{sym}(A)=\frac{A}{4(A-2Z)}\{\frac{1}{n(n+1)}\sum_{i=1}^{n}[B(A,Z+i)-B(A,Z-i)]+0.642\frac{2Z-1}{A^{1/3}}\}.
\end{eqnarray}
Figure 3 presents the experimental symmetry energy coefficient $a_{sym}(A)$ as a function of mass number for four different Coulomb energy expressions set I-IV. The dashed curve and the dashed-dotted curve denote the results with the Coulomb energy expressions $E_{c}=\frac{0.625}{A^{1/3}}Z^{2}$ and $E_{c}=\frac{0.642}{A^{1/3}}Z(Z-1)$, respectively, according to the Eqs. (9) and (14). The red solid curve and the black solid curve are the result of with the Coulomb energy expressions set II and set IV by adding the Coulomb correction terms to the Coulomb energy expression. From Fig.3 one can see that the values of the extracted symmetry energy coefficients increase by about 2 MeV for heavy nuclei when the Coulomb correction term is involved. Therefore the extraction of nuclear symmetry coefficient is dependence on the Coulomb energy expression and its coefficients.

\begin{figure}
\includegraphics[angle=-0,width= 0.65\textwidth]{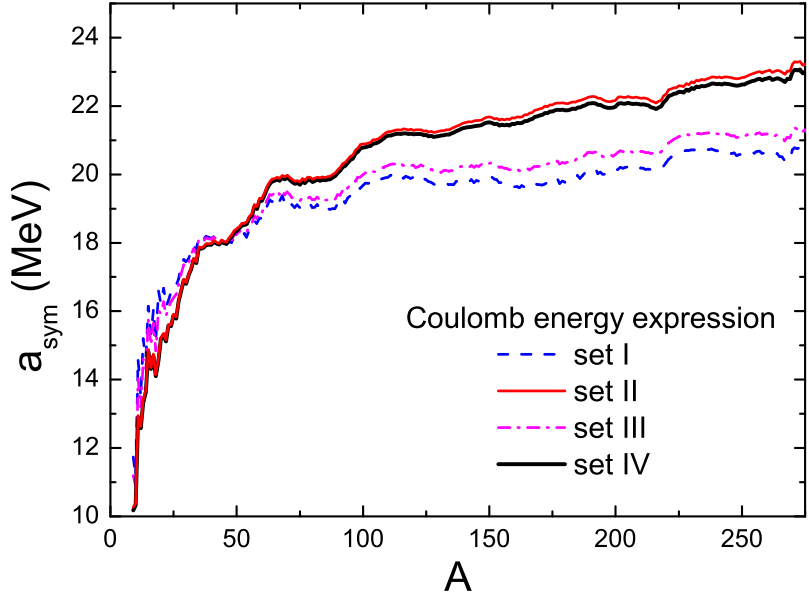}
 \caption{(Color online) Experimental symmetry energy coefficient $a_{sym}(A)$ as a function of mass number for four sets different Coulomb energy expressions.}
\end{figure}

The shell behavior is apparent in Fig. 2 in our approach by Eqs. (11) and (12). When the shell corrections are taken into account,
the fluctuations in the extracted $a_{sym}(A)$ are reduced effectively. However the reduction of fluctuations depends on the selecting shell correction energy.
Figure 4 shows the reduction of fluctuations of symmetry energy coefficient $a_{sym}(A)$ for three sets different shell correction energies of WS \cite{Wang10}, FRDM \cite{Moller95} and KTUY \cite{KTUY05}, respectively.
From Fig.4 one can seen that the effect of the shell correction energy from KTUY is so much smoother than that of the other two models, so the shell correction energies from KTUY is adopted in the calculation.
\begin{figure}
\includegraphics[angle=-0,width= 0.65\textwidth]{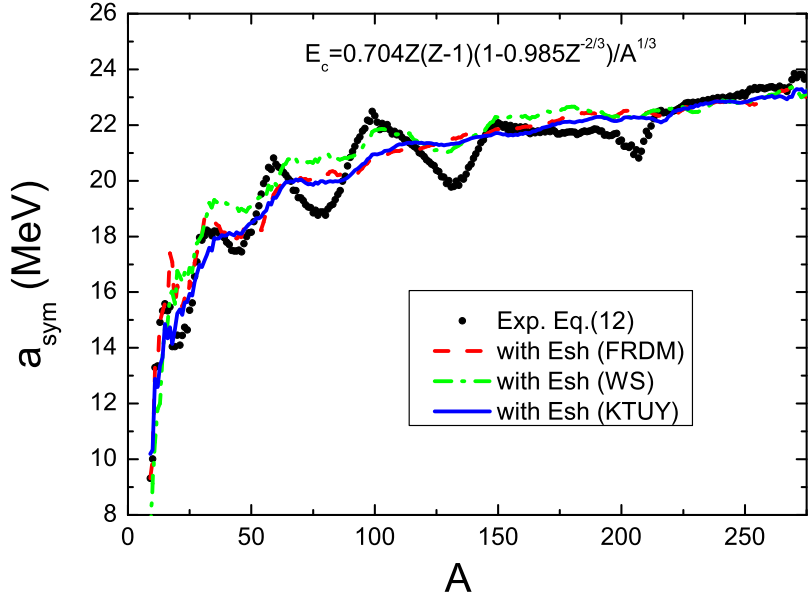}
 \caption{(Color online) The same as Fig.2 from Eq. (12), but with three different shell correction energies.}
\end{figure}

Figure 5 shows the experimental symmetry-energy coefficient $a_{sym}(A)$ as a function from Eq. (12) without the shell corrections (solid circles) and with the shell corrections energies of KTUY  (crosses).
The red curve and the green curve denote the fitting results of two definitions. By performing the two-parameter fitting to the $a_{sym}(A)$ obtained from Eq. (12) with the shell corrections of KTUY \cite{KTUY05} for 262 data, we evaluate $S_0=28.26\pm 1.3$ MeV and the surface-to-volume ratio $\kappa=1.26\pm 0.25$ MeV if assuming the mass dependence of symmetry energy coefficient $a_{\rm sym}(A)=S_0(1-\kappa/A^{1/3})$, and $S_0=32.80\pm 1.7$ MeV, $\kappa=2.82\pm 0.57$ MeV when $a_{\rm sym}(A)=S_0 (1+\kappa/A^{1/3})^{-1}$ is adopted. If we take the shell corrections of nuclei from WS \cite{Wang10} and the Coulomb energy $E_{c}=\frac{0.71}{A^{1/3}}Z^{2}(1-0.76Z^{-2/3})$, the results $S_{0}=31.1$ MeV and $\kappa=2.31$ MeV in Ref. \cite{Liu10} can be reproduced by using the formula $a_{\rm sym}(A)=S_0 (1+\kappa/A^{1/3})^{-1}$.

\begin{figure}
\includegraphics[angle=-0,width= 0.65\textwidth]{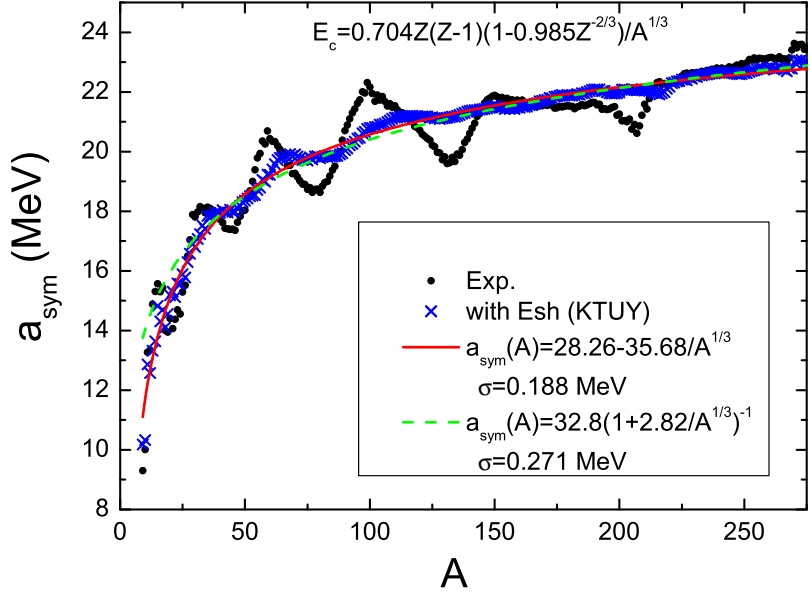}
 \caption{(Color online)  The same as Fig.4 from Eq. (12), but add the fitting results for two different definitions for the symmetry energy coefficient.}
\end{figure}

\section{\label{sec:level4}Summary}

In summary, we have proposed an alternative method to extract the symmetry energy coefficient of finite nuclei from the differences of available experimental binding energies of isobaric nuclei. In this approach, the influence of other effects can be effectively removed, except the Coulomb energy term. It is found that the Coulomb energy expression directly affects the value of the extracted symmetry energy coefficient. The Coulomb exchange correction plays an important role in the determination of the volume and surface symmetry coefficients. The symmetry energy coefficient increases by about 2 MeV for heavy nuclei when the Coulomb exchange term is involved. By performing a two-parameter fitting to the extracted $a_{sym}(A)$,  we obtain the bulk symmetry energy coefficient $S_0=28.26\pm1.3$ MeV and the surface-to-volume ratio $\kappa=1.26\pm 0.25$ MeV if assuming the mass dependence of symmetry energy coefficient $a_{\rm sym}(A)=S_0(1-\kappa/A^{1/3})$, and $S_0=32.80\pm1.7$ MeV, $\kappa=2.82\pm0.57$ MeV when $a_{\rm sym}(A)=S_0 (1+\kappa/A^{1/3})^{-1}$ is adopted. It indicates that the model dependence in the extraction of symmetry energy coefficient is strong and can not be ignored.

 \vspace{6pt}
This work was supported by National Natural Science Foundation of
China, Nos. 11005003, 11275052 and 11005002, the Natural Science Foundation of He'nan Educational Committee Nos.2011A140001 and 2011GGJS-147, and innovation fund of undergraduate at Anyang Normal University.

\end{document}